\documentclass[amsmath,amssymb,preprint,superscriptaddress,nobalancelastpage,aps,onecolumn,showpacs]{revtex4-1}
\usepackage{graphicx}
\usepackage[english]{babel}
\usepackage{amsmath}
\usepackage{amsfonts}
\usepackage{amssymb}
\usepackage[dvipsnames]{xcolor}

\usepackage{hyperref}
\hypersetup{colorlinks,linkcolor=blue,urlcolor=blue,citecolor=blue}

\usepackage{color}
\usepackage{ulem}

\begin{document}

\title{Link between Weyl-fermion chirality and spin texture}

\author {Kenta~Hagiwara}
\email{k.hagiwara@fz-juelich.de}
\affiliation {Peter Gr\"{u}nberg Institut (PGI-6), Forschungszentrum J\"{u}lich, J\"{u}lich 52425, Germany}
\affiliation {Fakult\"{a}t f\"{u}r Physik, Universit\"{a}t Duisburg-Essen, Duisburg 47057, Germany}

\author {Philipp~R\"{u}ßmann}
\affiliation{Institute for Theoretical Physics and Astrophysics, University of W\"{u}rzburg 97074, W\"{u}rzburg, Germany}
\affiliation {Peter Gr\"{u}nberg Institut (PGI-1) and Institute for Advanced Simulation, Forschungszentrum J\"{u}lich and JARA, J\"{u}lich 52425, Germany}

\author {Xin~Liang~Tan}
\affiliation {Peter Gr\"{u}nberg Institut (PGI-6), Forschungszentrum J\"{u}lich, J\"{u}lich 52425, Germany}
\affiliation {Fakult\"{a}t f\"{u}r Physik, Universit\"{a}t Duisburg-Essen, Duisburg 47057, Germany}

\author {Ying-Jiun~Chen}
\affiliation {Peter Gr\"{u}nberg Institut (PGI-6), Forschungszentrum J\"{u}lich, J\"{u}lich 52425, Germany}
\affiliation {Fakult\"{a}t f\"{u}r Physik, Universit\"{a}t Duisburg-Essen, Duisburg 47057, Germany}


\author {Keiji~Ueno}
\affiliation {Department of Chemistry, Saitama University, Saitama 338-8570, Japan}

\author {Vitaliy~Feyer}
\affiliation {Peter Gr\"{u}nberg Institut (PGI-6), Forschungszentrum J\"{u}lich, J\"{u}lich 52425, Germany}
\affiliation {Fakult\"{a}t f\"{u}r Physik, Universit\"{a}t Duisburg-Essen, Duisburg 47057, Germany}

\author {Giovanni~Zamborlini}
\affiliation {Peter Gr\"{u}nberg Institut (PGI-6), Forschungszentrum J\"{u}lich, J\"{u}lich 52425, Germany}

\author {Matteo~Jugovac}
\affiliation {Peter Gr\"{u}nberg Institut (PGI-6), Forschungszentrum J\"{u}lich, J\"{u}lich 52425, Germany}

\author {Shigemasa~Suga}
\affiliation {Peter Gr\"{u}nberg Institut (PGI-6), Forschungszentrum J\"{u}lich, J\"{u}lich 52425, Germany}
\affiliation {Institute of Scientific and Industrial Research (SANKEN), Osaka University, Osaka 567-0047, Japan}

\author {Stefan~Bl\"{u}gel}
\affiliation {Peter Gr\"{u}nberg Institut (PGI-1) and Institute for Advanced Simulation, Forschungszentrum J\"{u}lich and JARA, J\"{u}lich 52425, Germany}

\author {Claus~Michael~Schneider}
\affiliation {Peter Gr\"{u}nberg Institut (PGI-6), Forschungszentrum J\"{u}lich, J\"{u}lich 52425, Germany}
\affiliation {Fakult\"{a}t f\"{u}r Physik, Universit\"{a}t Duisburg-Essen, Duisburg 47057, Germany}
\affiliation{Department of Physics, University of California Davis, Davis CA 95616, USA}

\author {Christian~Tusche}
\email{c.tusche@fz-juelich.de}
\affiliation {Peter Gr\"{u}nberg Institut (PGI-6), Forschungszentrum J\"{u}lich, J\"{u}lich 52425, Germany}
\affiliation {Fakult\"{a}t f\"{u}r Physik, Universit\"{a}t Duisburg-Essen, Duisburg 47057, Germany}

\pacs{}    \textcolor[rgb]{0.00,0.07,1.00}{}

\maketitle

\textbf{Topological semimetals have recently attracted great attention due to prospective applications governed by their peculiar Fermi surfaces \cite{Armitage2018, Burkov2016}. Weyl semimetals host chiral fermions that manifest as pairs of non-degenerate massless Weyl points in their electronic structure,
giving rise to novel macroscopic quantum phenomena such as the chiral anomaly \cite{Jia2016}, an unusual magnetoresistance \cite{Ali2014, Chen2016}, and various kinds of Hall effects \cite{Zhang2018, Chen2018, Qian2014, Zhou2019, Song2020}. These properties enable the engineering of non-local electric transport devices, magnetic sensors and memories, and spintronics devices \cite{Yang2016, Kou2017}. Nevertheless, little is known about the underlying spin- and orbital-degrees of freedom of the electron wave functions in Weyl semimetals, that govern the electric transport.
Here, we give evidence that the chirality of the Weyl points in the Type-II Weyl semimetal MoTe$_2$ is directly linked to the spin texture and orbital angular momentum of the electron wave functions. By means of state-of-the-art spin- and momentum-resolved photoemission spectroscopy the spin- and orbital texture in the Fermi surface is directly resolved. Supported by first-principles calculations, we examined the relationship between the topological chiral charge and spin texture, which significantly contributes to the understanding of the electronic structure in topological quantum materials.}

Relativistic particles which attain a specific chirality, so-called Weyl fermions were proposed first in high-energy physics.
The signature of Weyl fermions, however, has still not been captured in high-energy physics, whereas Weyl fermions in condensed-matter systems have been discovered in topological semimetals \cite{Armitage2018, Burkov2016}.
Weyl fermions in solids emerge as massless band crossing points in the electronic structure, similar to Dirac fermions in topological quantum materials \cite{Hasan2010}, but always appear in pairs with opposite chirality and therefore need to be non-degenerate.
In general, topological quantum materials are classified according to the topology of their band structure.
The chirality of Weyl fermions can be characterized by a topological invariant, commonly called topological chiral charge.
These pairs of Weyl points (WPs) are protected by topology and are connected by Fermi arcs.
To realize such pairs of non-degenerate Weyl points, space-inversion or time-reversal symmetry needs to be broken in the system.

Topological chiral charges are related to the spin of the electrons as a fundamental quantum property.
Through spin-orbit coupling, the spin- and orbital-angular momentum are linked.
Thus, the spin- and orbital-degrees of freedom of the electron govern topological and transport properties.
First-principles calculation and spin-resolved photoemission studies suggest that Fermi arcs exhibit spin-momentum-locked spin textures in the type-I Weyl semimetals (Ta, Nb)(As, P) \cite{Sun2015,Lv2015, Xu2016} and the type-I\hspace{-.1em}I Weyl semimetal WTe$_{2}$ \cite{Feng2016}, similar to the chiral edges states of a topological insulator \cite{Hasan2010}.
Previous studies on the type-I\hspace{-.1em}I Weyl semimetal MoTe$_{2}$ have indeed reported the existence of spin-polarized bulk and trivial surface states \cite{Sakano2017, Crepaldi2017, Weber2018}, but not yet resolved the spin texture of the Weyl cones and the Fermi arc. 
In general, an experimental access to the spin texture of the topological states in Dirac- and Weyl semimetals is challenging due to the appearance of features in very small regions in energy- and momentum-space.
Furthermore, the WP location in 1$T_{d}$ MoTe$_{2}$ was previously only predicted theoretically and matter of the choice of simulation parameters \cite{Sun2015a, Chang2016, Wang2016, Ruessmann2018, Singth2020}.

Spin-resolved experiments for such materials with highest resolution have for a long time been hindered by the inefficiency of available photoemission spectroscopy tools. 
This situation has only recently been overcome by the advent of advanced electron spin analysers \cite{Tusche2015,Tusche2020,sug15,suga2021}.
Overall, the relationship between the spin texture and topological chiral charges in Weyl semimetals has still remained elusive.

In this work, we have performed detailed spin- and momentum-resolved photoemission studies by spin-resolving momentum microscopy on the space-inversion-broken type-I\hspace{-.1em}I Weyl semimetal 1$T_{d}$ MoTe$_{2}$ \cite{Clarke1978}.
In particular, by exploiting circular dichroism, we obtained a direct fingerprint of the chiral quasiparticle states.
Furthermore, the measured spin-resolved Fermi surface revealed the spin texture of the Weyl cone, which is linked to the chiral charge of the respective WPs.

\begin{figure}[t]
\includegraphics[width=8cm]{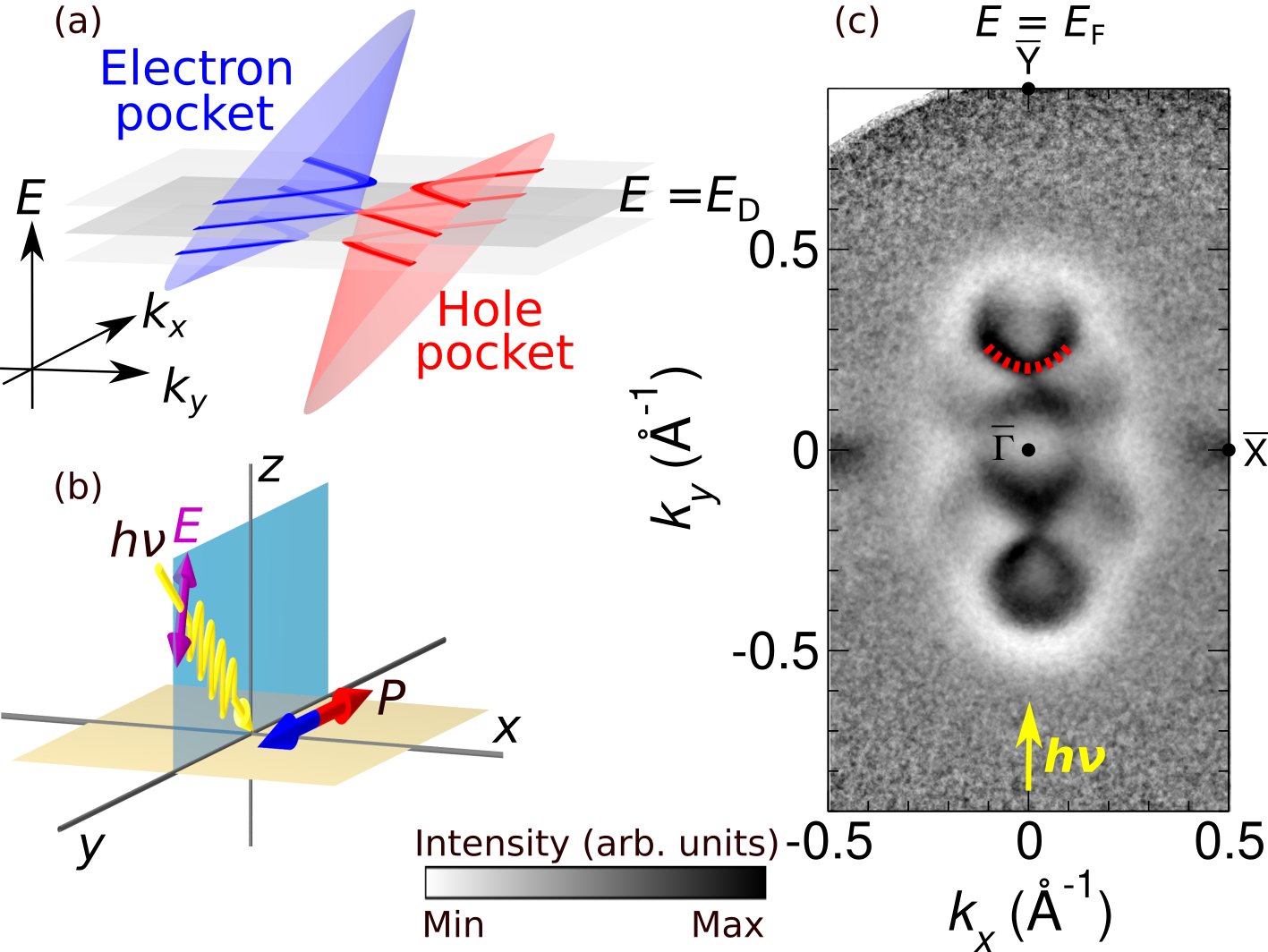}
\caption{
\textbf{Type-I\hspace{-.1em}I Weyl semimetal MoTe$_{2}$.}
(a) Strongly tilted cone characterizing the type-I\hspace{-.1em}I Weyl semimetal.
Weyl point appears at the boundary between electron and hole pockets at energy $E=E_\text{D}$.
(b) Photoemission experimental geometry.
A purple arrow indicates electric field ($E$) vector for p-polarized light. 
Detection of spin polarization ($P$), indicated by red and blue arrows, is along the $y$ direction in the light incidence plane for spin-resolved measurement.
(c) Measured Fermi-surface contour using p-polarized light at a photon energy of $h\nu$ = 52 eV.
High symmetry points of the surface Brillouin zone are indicated by the corresponding labels.
Red dotted line indicates the Fermi arc as a guide to the eye.
}
\label{fig:1}
\end{figure}

Type-I\hspace{-.1em}I Weyl semimetals are characterized by strongly tilted Dirac cones \cite{Soluyanov2015} as shown in Fig.~\ref{fig:1}a.
In the constant energy momentum map at the Fermi level displayed as a middle planar cut, the WPs appear at the boundary between electron and hole pockets forming the upper and lower half of the Dirac cones, respectively.
Away from the Fermi level, these electron- and hole-pockets open and are separated in the constant energy cut.

The photoemission experiments were performed in the geometry shown in Fig.~\ref{fig:1}b. 
The photon beam is incident in the $y$-$z$ plane at an angle of 25$^\circ$ with respect to the sample surface. 
The optical plane coincides with the $\overline{\Gamma}$-$\overline{\text{Y}}$ direction of the MoTe$_2$ surface Brillouin zone (BZ) as indicated in Fig.~\ref{fig:1}b. For the spin-resolved measurements discussed below, the spin quantization axis $\textbf{\textit{P}}$ points in-plane along the positive $k_\mathrm{y}$ axis.

Figure \ref{fig:1}c shows the measured Fermi-surface (FS) contour using p-polarized light at a photon energy of $h\nu$ = 52 eV.
We have performed measurements at $T \sim$ 100 K, where MoTe$_{2}$ crystallizes in the 1$T_{d}$ structure \cite{Clarke1978} and becomes a type-I\hspace{-.1em}I Weyl semimetal.
One can clearly see the hole-pocket centred at the BZ centre and two electron-pockets around the positive and negative $k_y \approx\pm0.3\,\mathrm{\AA}^{-1}$ at $k_x = 0$. These electron- and hole-pockets form the tilted Dirac cones in the upper and lower half of the MoTe$_2$ FS contour. The measured FS section here corresponds to the planar cut displayed in Fig.~\ref{fig:1}a, such that the Dirac cone states appear with a linear crossing in the constant energy photoelectron momentum map.
The observed band dispersions are overall consistent with previous reports \cite{Ruessmann2018, Sakano2017, Crepaldi2017, Weber2018, Sun2015a, Wang2016, Chang2016, Deng2016, Tamai2016, Jiang2017}.

One can also observe the Fermi arc being located outside of the electron pocket toward the touching points between the electron and hole pockets, as indicated in Fig.~\ref{fig:1}c by red dotted lines as a guide to the eye. This result is in good agreement with previous observations for this material \cite{Tamai2016, Weber2018}. 
While the end points of the arc are supposed to connect to WPs with opposite chiral charge, a direct observation of the WPs in  1$T_{d}$ MoTe$_2$ is still lacking.

A fingerprint of the Weyl points and the associated chiral Dirac states can be obtained through the excitation by circularly polarized light in a photoemission experiment \cite{Park2012, Uenzelmann2021}. Interaction with circularly polarized light, which possesses a specific chirality itself, in general reflects the chirality of matter. A different response to left and right circularly polarized light, an effect commonly called ``circular dichroism'', can arise in non-magnetic materials due to a chiral crystal structure \cite{Hecht1974} or a chiral experimental geometry \cite{Schoenhense1990}. In Weyl semimetals, it has been shown that the correlation between circularly polarized light and the intrinsic chirality of the Weyl states drives an unidirectional photocurrent in response to the respective light helicity \cite{Chan2017, Ma2017}. In a photoemission experiment, the observation of the circular dichroism therefore provides a probe for identifying the Weyl points and 
their locations in energy and momentum space. 

\begin{figure*}[tb]
\includegraphics[width=17cm]{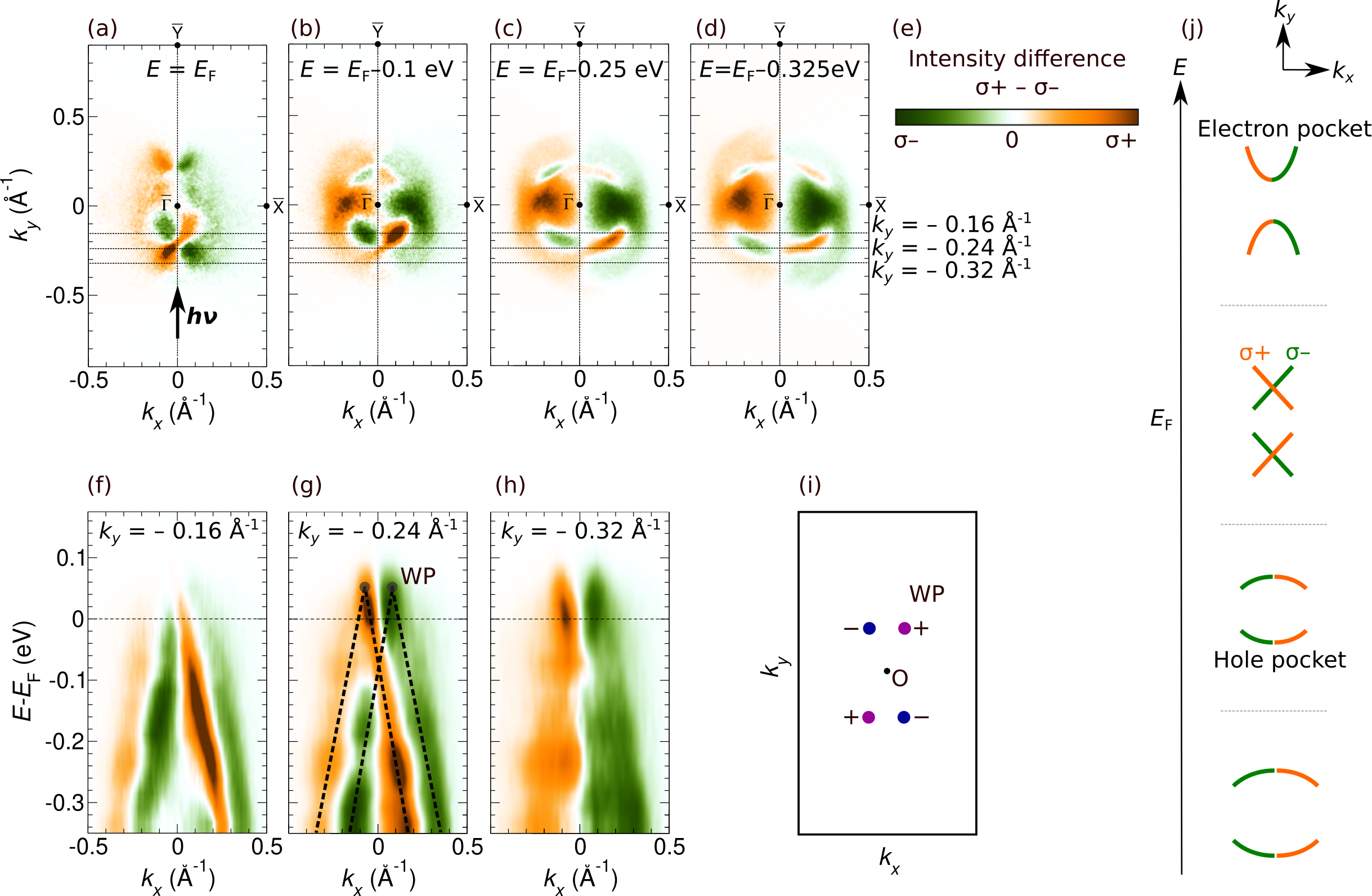}
\caption{
\textbf{Circular dichroism of the Weyl cone.}
(a-d) Constant-energy maps measured at a photon energy of $h\nu$ = 60 eV at $E = E_\text{F}$ (a), $E_\text{F}- 0.1\,$eV (b), $E_\text{F}- 0.25$\,eV (c), $E_\text{F}- 0.325\,$eV (d).
(f-h) Corresponding band-dispersion maps at $k_y = -0.16\,\mathrm{\AA}^{-1}$ (f), $-0.24\,\mathrm{\AA}^{-1}$ (g), $-0.32\,\mathrm{\AA}^{-1}$ (h).
Black dotted lines in panel (g) indicates the Weyl cones as a guide to the eye.
Plotted intensities indicate difference between right- ($\sigma+$) and left- ($\sigma-$) circularly polarized light according to the color code in panel (e).
(i) Configuration of the WPs in momentum space.
Purple and dark blue colors indicate the opposite chirality of the WPs. 
(j) Schematic band-structure development as a function of binding energy. 
}
\label{fig:2}
\end{figure*}

Figure \ref{fig:2} shows the circular dichroism in the angular distribution (CDAD) measured at a photon energy of $h\nu$ = 60 eV.
The features related to the Weyl cones depend on the photon energy, which is evident for the bulk Dirac cone.
Here, we chose the photon energy such that the section probed by the photon energy through the bulk BZ corresponds to the high-symmetry point $k_z=0$ and coincides with the bulk WP (for details see Methods).
The dichroism signal can be taken from the photoemission intensity difference between two measurements taken with right- and left-circularly polarized light. 
Orange and green intensities correspond to right- ($\sigma+$) and left- ($\sigma-$) circularly polarized light. 
In the Fermi surface contour in Fig.~\ref{fig:2}a, one can clearly see the CDAD texture of two ``X"-shaped crossings originating from the boundary between the hole pocket and two electron pocket.
According to the measured constant-energy maps in Figs.~\ref{fig:2}b-d, the hole pocket becomes larger towards larger binding energies.

A schematic evolution of the corresponding band features as a function of the binding energy is summarized in Fig.~\ref{fig:2}j.
Electron pockets are located above $E_\text{F}$ (top). Moving down in energy towards $E_\text{F}$, electron and hole pockets touch each other, corresponding to the crossings (second schematic from top) that we observed in Fig.~\ref{fig:2}a. At larger binding energies the crossing point again opens and the constant energy contour shows the hole pockets.

Circularly polarized light couples not only to an intrinsic chiral system, but also to a chirality that is induced by the handedness of the experimental setup \cite{Schoenhense1990}.
In more detail, the incident photon vector \textit{\textbf{q}}, the photoelectron momentum \textit{\textbf{p}}, and the surface normal \textit{\textbf{n}} define a handed coordinate system.
We therefore chose an experimental geometry where the incident beam is perpendicular to a crystal mirror plane. A CDAD that might arise from the experimental geometry then vanishes at the $k_x=0$ line, where all three vectors \textit{\textbf{q}}, \textit{\textbf{p}}, and \textit{\textbf{n}} lie in the same plane, and no handedness is defined.
The predicted WPs of 1$T_{d}$ MoTe$_{2}$ are located close to $k_x=0$. Here, the geometry contribution to the CDAD minimizes and the dichroic signal strongly reflects the chirality of the electronic structure. 
Therefore, this experimental geometry is best suited to get access to the chiral charge of the Weyl points.

Figures~\ref{fig:2}f-h show the band dispersion plotted along cuts parallel to the $\overline{\text{X}}-\overline{\Gamma}-\overline{\text{X}}$ direction (see horizontal dotted lines in Figs.~\ref{fig:2}a-d). 
The CDAD here reveals more details of the development of the Weyl cones.
In particular, we observe that the pair of Weyl cones exhibits a strong CDAD signal, whereas the sign is reversed between the two cones, following the expectation of opposite chiral charges that these states carry. The CDAD fingerprint allows one to determine the Weyl point energy to be located +50$\,$meV above $E_\mathrm{F}$.

Figure \ref{fig:2}i shows the configuration of the WPs in momentum space \cite{Ruessmann2018}.
In a Weyl semimetal with time-reversal symmetry, but without space inversion-symmetry, such as 1$T_{d}$ MoTe$_{2}$, the WPs distribute symmetrically in momentum ($k$) space.
The total number of Weyl points must be a multiple of four \cite{Armitage2018, Hirayama2018}, because under time reversal operation the WP at the momentum $k_0=(k_{x_0}, k_{y_0}, k_{z_0})$ is converted into the WP at $-k_0=(-k_{x_0}, -k_{y_0}, -k_{z_0})$ with the same chirality.
In 1$T_{d}$ MoTe$_{2}$, the WPs are located at $k_z=0$.
Thus, the WP at ($k_{x_0}$, $k_{y_0}$) denoted by ($+$) and the WP at ($-k_{x_0}$, $k_{y_0}$) denoted by ($-$) have opposite chirality.

The Weyl phase associated with the WPs in MoTe$_{2}$ and Mo$_{x}$W$_{1-x}$Te$_{2}$ has been debated and is still controversial \cite{Ruessmann2018, Sakano2017, Crepaldi2017, Sun2015a, Wang2016, Chang2016, Deng2016, Tamai2016, Jiang2017}.
According to theoretical studies, the number of WPs depends on lattice constant, strength of spin-orbit coupling, and atom positions \cite{Chang2016, Ruessmann2018}.
From our experimental results we could identify 4 WPs located slightly above the Fermi level (our experimental value $\sim E_{\textrm{F}}+50$ meV).
Based on our results, we resolved the long-standing question of where the WPs are located.
This is important because the physical and transport properties are dominated by the Weyl phase near the Fermi level.
Therefore, MoTe$_2$, where the WPs are located near the Fermi level, exhibits, in particular, spin-dependent transport  properties \cite{Zhou2019, Song2020}, and is a candidate for realizing Weyl-band spintronics applications mentioned above \cite{Yang2016, Kou2017}.

\begin{figure}[t]
\includegraphics[width=8cm]{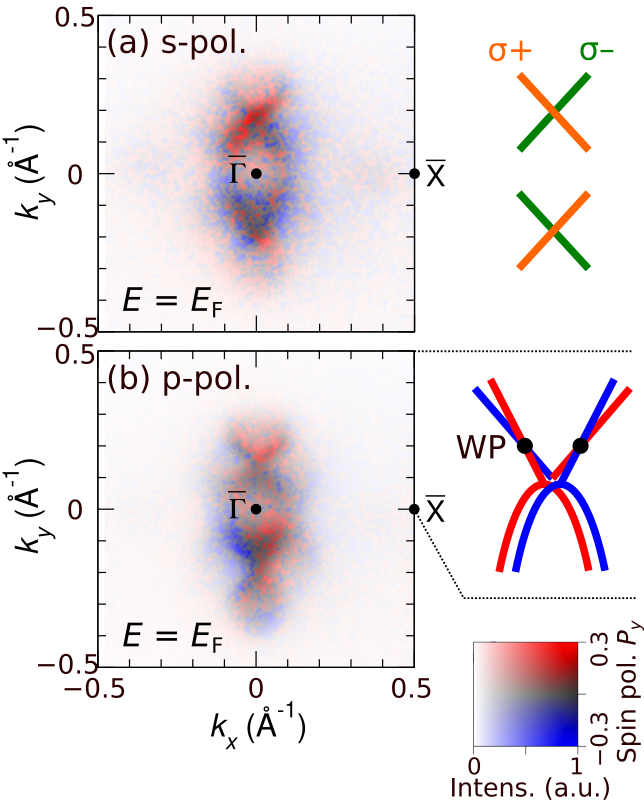}
\caption{
\textbf{Spin texture of the Weyl cone.}
Spin-resolved Fermi-surface contour measured using s-(a) and p-(b) polarized light at a photon energy $h\nu$ = 52 eV.
Sketches on the right summarize the observed Fermi surface features in $(k_x, k_y)$ as a guide to the eye.
A sketch in panel (a) shows the observed CDAD texture from Fig.~\ref{fig:2}a by using the same green and orange color code as in Fig.~\ref{fig:2}.
In panel (b), a sketch is shown in the $k_y>0$ upper half region.
Spin-resolved intensities are encoded using the 2D colour code: red and blue intensities indicate spin-up and spin-down photoelectrons
with a spin quantization axis along the $k_\mathrm{y}$ direction.
}
\label{fig:3}
\end{figure}

Figure \ref{fig:3} shows the spin-resolved Fermi surface contour measured at a photon energy of $h\nu$ = 52 eV.
Note that the FS contour depends only little on the photon energy.
This is consistent with the calculated Fermi surface in Ref.~\citep{Weber2018}, showing a weak $k_z$ dependence of the FS contour.
This is because the WPs are located slightly above the Fermi level and the shape of the Fermi surface is not sensitive to small change in photon energy.
Using s-polarized light (Fig.~\ref{fig:3}a), the observed spin texture shows a single crossing of spin-up (red) and spin-down (blue). The measured spin polarization is always given with respect to the positive $k_\mathrm{y}$ axis.
This spin texture corresponds well to the observed CDAD texture from Fig.~\ref{fig:2}a, which is here also displayed on the right hand side of Fig.~\ref{fig:3}a as a reference.
For p-polarized light, shown in Fig.~\ref{fig:3}b, the observed spin texture shows additional details due to the different orbital selectivity. In particular, the ''X"-shaped contours of the Dirac cone states appear as double lines, where a spin-down (blue) state is located slightly left of a spin-up (red) state. The sketch on the right hand side of Fig.~\ref{fig:3}b summarizes the observed FS contour in the $k_y>0$ region as a guide to the eye. The $k_{||}$ separation of the spin-split states that we observe here is too small to be resolved in a spin-integrated measurement, e.\,g., in Fig.~\ref{fig:1}.

The spin splitting of the Weyl cone states is caused by a combination of spin-orbit coupling and the broken inversion symmetry. In particular, the observed reversed spin polarization to the two cones originates from an opposite chirality of the respective WPs.
Our observation of different spin textures for measurements taken with p-polarized light versus s-polarized light, where no spin-splitting of the ''X"-shaped contour is observed, can be explained by optical selection rules that  p- and s- polarized light selectively probe contributions to the Weyl cone states of even and odd orbitals, respectively.
This additionally gives rise to an asymmetry of photoemission intensities and spin polarization between the top and bottom half of the Fermi surface image, in reminiscence of the spin resolved photoemission observations on topological insulators and related materials \cite{Meyerheim2018, maa16}.
The latter effect stems from the tilted electric field vector for p-polarized light (Fig.~\ref{fig:1}b) that breaks the mirror symmetry along the $x$-$z$ plane, an effect known as linear dichroism in the angular distribution (LDAD).

\begin{figure}[t]
\includegraphics[width=8cm]{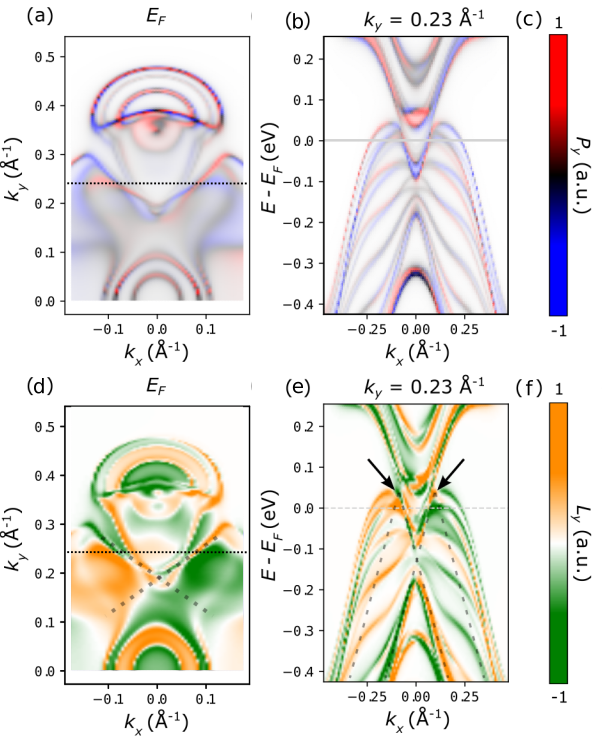}
\caption{
\textbf{Calculated spin polarization and orbital angular momentum.} 
The spin polarization $P_y$ (a,b) and the orbital angular momentum $L_y$ (d,e) is integrated over the middle two MoTe$_2$ layers of an eight layer thick MoTe$_2$ film.
The color scales of $P_y$ and $L_y$ are given in (c) and (f), respectively.
The dashed grey lines in (d,e) and the arrows serve as guides to the eye and indicate the position of the Weyl points WP.
}
\label{fig:4}
\end{figure}

Figures \ref{fig:4}a and \ref{fig:4}b show the calculated spin-polarized Fermi surface contour and the band-dispersion map cutting through the WPs, respectively.
In agreement with our experimental observation, 4 WPs exist in the calculation as discussed in detail in Ref.~\cite{Ruessmann2018}.
The calculated spin polarization in Fig.~\ref{fig:4} indicates bulk-like character integrated over the middle 2 layers of a 8-layer thick  MoTe$_2$ film to account for the more bulk-sensitivity of our measurement. As in the experiments, the spin polarization along the $k_\mathrm{y}$ direction is shown.
Careful inspection of the boundary between the electron- and hole-pockets in the calculated Fermi surface (Fig.~\ref{fig:4}a) reveals a spin texture similar to our experimental result excited by p-polarized light in Fig. \ref{fig:3}b.
The good agreement of the measurements with the calculation clarified that the experiment mainly probes bulk-like states.
Furthermore, our calculation confirms the spin-splitting of the Weyl cone states, that we have observed by p-polarized light.

For a deeper understanding of the observed circular dichroism we have also calculated the orbital angular momentum $L_\mathrm{y}$ in the Fermi surface, shown in Figs.~\ref{fig:4}d and \ref{fig:4}e.
It has been shown that the CDAD probes the orbital angular momentum of the wave function \cite{Schoenhense1990}.
The CDAD can be shown to be approximately proportional to the projection of the orbital angular momentum \textit{\textbf{L}} on the light propagation direction \cite{Park2012, Uenzelmann2021}.
A recent study further demonstrated that momentum mapping of the orbital angular momentum reflects the chirality of the WPs in a Weyl semimetal \cite{Uenzelmann2021}.
In our experimental geometry, where the light incidence is aligned in the $y$-$z$ plane, the observed CDAD is sensitive to the $L_{y}$ component of the orbital angular momentum.

The orbital momentum of the wavefunctions at the Fermi level in Fig.~\ref{fig:4}d shows a rich texture with sign changes at the positions of the Weyl points. This shows the correlation between the chirality of the Weyl points and the orbital texture of the electronic structure around them \cite{Uenzelmann2021}. Considering possible matrix element effects we find a good agreement of the calculated values with the measurements, that show a crossing of the changing chirality around $k_y=0.2\,\mathrm{\AA}^{-1}$ (highlighted by dotted lines). 
Figure \ref{fig:4}e shows an energy-momentum section along the $\overline{\Gamma}-\overline{\text{X}}$ direction.
The dispersion of the measured circular dichroism features near the WPs (indicated by dashed lines in Fig.~\ref{fig:2}g) agrees well with the calculated $L_\mathrm{y}$ values (see dashed lines in Fig.~\ref{fig:4}e). This observation further supports our conclusion that strong circular dichroism with reversed sign reflects the opposite chiral charge.
As shown in Fig.~\ref{fig:3}a, the spin texture probed by s-polarized light corresponds well to the observed CDAD texture.
This suggests that the $P_y$ spin component of odd orbitals, probed by s-polarized light, couples to the $L_y$ orbital angular momentum probed by the CDAD.
We can also conclude that the observed spin texture of even orbitals, probed by p-polarized light (Fig.~\ref{fig:3}b), closely corresponds to the ground state spin texture of MoTe$_2$ calculated in Fig.~\ref{fig:4}a.

In summary, using different polarization states of the exciting light, as well as advanced first-principles calculations, we clarified the spin- and orbital-textures of the topological states.
This revealed fingerprints of the topological character of the quasiparticle states in reminiscence of previous results for type-I Weyl semimetals \cite{Uenzelmann2021}.
We captured a pair of WPs exhibiting a strong CD with reversed sign, giving evidence for the location of the WPs in energy-momentum space near the Fermi level and their respective chirality.
This is further supported by the DFT calculated map of the orbital angular momentum \textit{\textbf{L}} in the Fermi surface.
Our measured and calculated spin-resolved Fermi surface contours reveal, in good agreement, two closely located Weyl cones with reversed spin polarization, suggesting an opposite chirality of the respective WPs.
Our results clarify the topological electronic structure and location of the WPs in the Type II Weyl semimetal 1$T_{d}$ MoTe$_2$, and can further be applied not only to Weyl semimetals, but also to the deeper understanding of generic topological quantum materials.

\section{Methods}

High-quality single crystals of MoTe$_{2}$ were synthesized as described in Ref \cite{Ueno2015}. Spin- and momentum-resolved photoemission experiments were performed at the NanoESCA beamline \cite{Wiemann2011} of the Elettra synchrotron in Trieste (Italy), using linearly and circularly polarized light in the energy range between 35eV and 60eV.
All measurements were performed while keeping the sample at $T \sim$ 100 K, where MoTe$_{2}$ crystallizes in the 1$T_{d}$ structure.
Photoelectrons emitted into the complete solid angle above the sample surface were collected using a momentum microscope \cite{Tusche2019}.
Spin-resolved photoelectron maps were measured using a W(100) imaging spin filter combined with momentum microscopy \cite{Tusche2011, Tusche2013, Tusche2020, suga2021}.
Spin polarization ($P$) is detected along the $k_\mathrm{y}$ in-plane direction in the light incidence plane, as shown in Fig.~\ref{fig:1}b.
The crystal was cleaved \textit{in situ} at room temperature and measured in an ultrahigh vacuum of better than $\sim$ $10^{-10}$ Torr.

The photon energy $h\nu$ is converted to $k_z$ via $k_z=\sqrt{\frac{2m}{\hbar^2}(h\nu-\phi+U_i)-{k_{\parallel}}^2}$.
Here, we determined an inner potential $U_i=7 \text{eV}+\phi$ ($\phi$: Work function of the analyzer) experimentally such that $h\nu=60\,$eV corresponds to the high-symmetry point $k_z=0$.
A tomographic section through the BZ can be probed by momentum microscopy and the corresponding $k_z$ can be selected by the photon energy, as described in detail in Ref.~\citep{Tusche2018}.

Density functional theory (DFT) calculations were carried out for an eight layer thick film of MoTe$_2$ within the local spin density approximation \citep{Vosko1980} using the full-potential relativistic Korringa-Kohn-Rostoker Green's function method (KKR) \citep{Ebert2011,jukkr} with exact description of the atomic cells \citep{Stefanou1990,Stefanou1991}. The truncation error arising from an $\ell_{max} = 3$ cutoff in the angular momentum expansion was corrected for using Lloyd's formula \citep{Zeller2004}. Further details on the calculation setup can be found in \cite{Weber2018}.
The expectation values for the spin and orbital angular momentum operators were calculated from the Green's function using $A_i = -\frac{1}{\pi}\mathrm{Im} \int\limits_{-\infty}^{E_F} \mathrm{Tr}\left[\hat{A}_i\,G(E)\right] \mathrm{d}E$ where $\hat{A}_i = \hat{\sigma}_i, \hat{L}_i $ and $i=x,y,z$. They are plotted in terms of their spectral density in Fig.~\ref{fig:4}. We chose to show the spin polarization and the orbital angular momentum on the Fermi surface integrated over the central two MoTe$_2$ layers of the eight layer film to account for the more bulk-like sensitivity of the photoemission experiments.


\section{Acknowledgments}
Photoemission experiments were performed at Elettra synchrotron in Trieste, Italy (Proposal Nos. 20180023 and 2018053).
We acknowledge K.\ Fukushima (Saitama University, Japan) for supporting the synthesis and characterization of the samples.
We acknowledge computing time granted by the JARA Vergabegremium and provided on the JARA Partition part of the supercomputer CLAIX at RWTH Aachen University.
We thank the Bavarian Ministry of Economic Affairs, Regional Development and Energy for financial support within High-Tech Agenda Project ``Bausteine f\"{u}r das Quantencomputing auf Basis topologischer Materialien mit experimentellen und theoretischen Ans\"{a}tze''.
This work was funded by the BMBF under Grant No.~05K19PGA and by the Deutsche Forschungsgemeinschaft (DFG, German Research Foundation) under Germany's Excellence Strategy – Cluster of Excellence Matter and Light for Quantum Computing (ML4Q) EXC 2004/1 – 390534769.


\section{Author contributions}

KH, XLT, YJC, VF, SS, CT performed the experiments with assistance by GZ, MJ.
KH analyzed the experimental data with suggestions by XLT, YJC, SS, and supervised by CT.
PR performed calculations under supervision of SB.
KU synthesized and characterized the samples.
KH, PR, CT drafted the manuscript.
CT designed and coordinated the research together with SS, CMS. All authors discussed the results and contributed to the manuscript.


%

\newpage

\end{document}